\documentclass[aps,pra,nofootinbib,amsmath,amssymb,amsfonts,superscriptaddress,floatfix]{revtex4} 
\pdfoutput=1
\usepackage{amsmath,amsthm,amsfonts,amssymb}
\usepackage{graphicx}

\usepackage{color}

\newtheorem{lemma}{Lemma}

\newcommand{\ocite}{\cite}

\DeclareMathAlphabet{\mathpzc}{OT1}{pzc}{m}{it}

\newcommand{\be}{\begin{equation}}
\newcommand{\ee}{\end{equation}}

\newcommand{\expec}{\mathbb{E}}
\newcommand{\pr}{{\rm Pr}}
\newcommand{\ze}{{\hat Z}}
\newcommand{\yi}{{\hat Y}}
\newcommand{\ex}{{\hat X}}
\newcommand{\Val}{{\rm Val}}
\newcommand{\Cut}{{\rm Cut}}

\begin{document}

\title{Classical and Quantum Bounded Depth Approximation Algorithms}

\author{Matthew B.~Hastings}

\affiliation{Station Q, Microsoft Research, Santa Barbara, CA 93106-6105, USA}
\affiliation{Quantum Architectures and Computation Group, Microsoft Research, Redmond, WA 98052, USA}
\begin{abstract}
We consider some classical and quantum approximate optimization algorithms with bounded depth.  First, we define a class of ``local" classical optimization algorithms and show that a single step version of these algorithms can achieve the same performance as the single step QAOA on MAX-3-LIN-2.  Second, we show that this class of classical algorithms generalizes a class previously considered in the literature\ocite{hirvonen2014large}, and also that a single step of the classical algorithm will outperform the single-step QAOA on all triangle-free MAX-CUT instances.  In fact, for all but $4$ choices of degree, existing single-step classical algorithms already outperform the QAOA on these graphs, while for the remaining $4$ choices we show that the generalization here outperforms it.
Finally, we consider the QAOA and provide strong evidence that, for any fixed number of steps, its performance on MAX-3-LIN-2 on bounded degree graphs cannot achieve the same scaling as can be done by a class of ``global" classical algorithms.  These results suggest that such local classical algorithms are likely to be at least as promising as the QAOA for approximate optimization.
\end{abstract}
\maketitle

This paper considers various approximate optimization algorithms based on bounded depth circuits, both classical and quantum.
One such quantum algorithm is the QAOA  (quantum approximate optimization algorithm)\ocite{farhi6062quantum}, while classical algorithms
with this property had been considered previously\ocite{hirvonen2014large,suomela2013survey}.

Although initially the QAOA generated much excitement due to it outperforming\ocite{farhi6062quantum} the best classical algorithm at the time on MAX-3-LIN-2, soon after a classical algorithm was constructed with asymptotically much better performance\ocite{barak2015beating}.
The one-step QAOA improved upon a random solution by an amount proportional to $\Theta(1/D^{3/4})$, where $D$ is the degree of the instance, while the classical algorithm attained an improvement $\Omega(1/D^{1/2})$; indeed, the classical algorithm attained this performance for MAX-K-LIN-2 for any odd $K$.

The one-step QAOA used, however, has the property that it can be done with bounded depth and without any nonlocal communication, in that one can initialize a product state, apply some bounded depth circuit, and produce a final state that obtains some given expectation value for the objective function, while the algorithm of Ref.~\onlinecite{barak2015beating} does not have this property.  
That is, using terminology we will define later, the QAOA is an example of a ``local" algorithm while the algorithm of Ref.~\onlinecite{barak2015beating} is not a local algorithm.  Such local algorithms have been well-studied in the classical setting\cite{suomela2013survey}.

In a second arXiv version of 
\ocite{farhi6062quantum}, it was shown that the performance of the QAOA could be improved to give a $\Omega(1/(D^{1/2} \ln(D)))$ improvement over random for MAX-3-LIN-2, which is still worse than the classical algorithm but better than the original QAOA.  This modification of the QAOA, however, gave an algorithm that is not local because it required making an appropriate global choice of some parameter (an angle in the QAOA).  This global choice must be done independently for each instance, so that some nonlocal communication is required.  Further, once we start considering algorithms that combine a bounded depth circuit with the choice of a small number of global parameters, then there is a very simple classical algorithm\ocite{qqdq} that attains the same performance $\Omega(1/D^{1/2})$ as the nonlocal classical algorithm of Ref.~\onlinecite{barak2015beating} while also having some additional guarantees (this classical algorithm\ocite{qqdq} can be explained in one sentence as: take the local algorithm that we give here in section \ref{M3L2}, globally optimize over parameter $c_0$ in an interval to maximize the absolute value of the objective function, and follow with a change in sign of spins if the objective function has the wrong sign).
In this paper, then, when referring to QAOA or other algorithms based on local circuits, we will not consider such algorithms with a choice of global parameters.

In this paper we further explore such local classical algorithms and we show that they outperform the QAOA for a variety of problems.
We define a class of such algorithms called local tensor algorithms.
We will explain later that the algorithm of Ref.~\onlinecite{hirvonen2014large} can be regarded as an example of these algorithms.
This allows us to generalize that algorithm to show that it outperforms the one-step QAOA for MAX-CUT on triangle-free graphs for all choices of degree; the classical algorithm is still local (and indeed is still a single step).

Another result of this paper is to expose a limitation of such local algorithms, especially in the particular case of QAOA, showing that for some problems, increasing the depth of the algorithm to any bounded amount will not lead to significant improvements in performance.

To fix notation, we will consider a variety of optimization problems of the form MAX-K-LIN-2 and MAX-CUT.  We will consider a problem with $N$ variables.  We call these variables ``spins", and we denote the value of the $i$-th variable by $Z_i$ with $Z_i\in \{-1,+1\}$.
MAX-K-LIN-2 is a problem where the objective function is a sum of monomials, each of which is a product of $K$ distinct spins.  For example for $K=2$ the objective function is
$\frac{1}{2}\sum_{i,j} J_{i,j} Z_i Z_j$ for some matrix $J_{i,j}$ that we will take to be symmetric, while for $K=3$ the objective function is
$\frac{1}{3!}\sum_{i,j,k} J_{i,j,k} Z_i Z_j Z_k$ for some tensor $J_{i,j,k}$ that we take to be symmetric in its indices (and to vanish if any indices coincide with each other, i.e., $J_{iij}=0$).  The prefactor $1/K!$ is a convenient normalization so that we take the tensors $J$ symmetric in their indices.

In many of these problems, one restricts to the case that the entries of $J$ are chosen from the set $\{-1,0,+1\}$.  MAX-CUT is, up to a linear transformation, an instance of this with $K=2$.  The case $K=2$ is the same as the Ising model in physics, up to an overall sign difference because we choose to maximize the objective function rather than minimize the energy.

For such a problem, an instance has degree $D$ if for every spin, that spin participates in $D$ distinct nonzero terms in the objective function.

\section{Local Algorithms}
We now define a class of ``local"  algorithms.  Our focus in this paper is on local classical algorithms, but we define local quantum algorithms for completeness.  In its simplest form, local classical algorithms include the algorithm of Ref.~\onlinecite{hirvonen2014large}, but we also consider generalizations.

We will call an algorithm a local algorithm if it is as follows.  We associate some number of additional degrees of freedom with each spin (the degree of freedom may be arbitrary, it need not be $\{-1,+1\}$.  Initially, these degrees of freedom are chosen from some random distribution, with the degrees of freedom corresponding to different spins chosen independently and from the same distribution.
Then, a sequence of steps is applied; in each step, the degrees of freedom on each spin are updated in some way that depends on the degrees of freedom on neighboring spins, where a pair of spins are said to be neighbors if they both participate in some term in the objective function.  Finally, after a bounded number of steps, the degrees of freedom are used to produce some configuration of spins, so that the value of the $i$-th spin $Z_i$ depends only on the degrees of freedom associated with that spin.

In a local classical algorithm, these degrees of freedom are classical variables and the allowed transformations are any classical computation.  Such classical algorithms are well-studied\cite{suomela2013survey}.  In a local quantum algorithm some degrees of freedom may be quantum, and the update is by a local quantum circuit (more generally, by completely-positive trace preserving maps, but we may add ancillary degrees of freedom so that all computations may be done with circuits).  The QAOA is an example of such a local quantum algorithm.

We will call the particular class of local algorithms studied in this paper ``tensor algorithms", for reasons that become more clear later.
As a simplest case, consider defining a vector of real numbers, with a one-to-one correspondence between entries of the vector and spins.  Initialize this vector in some random way, choosing each entry independently from some given distribution; below we will consider the case that this distribution is either the uniform distribution on $\{-1,+1\}$ or is the uniform distribution on the interval $[-1,+1]$.

Then, apply a series of steps; each step consists of an update to the vector depending on the given objective function, followed by some nonlinear operation acting on each entry of the vector independently.  After some number of steps, the entries of the vector are used to determine a final spin configuration.  This could be done in a variety of ways, for example either choosing the spin to equal $+1$ if the sign of the entry is positive and $-1$ if the sign is negative, choosing it randomly if the sign is zero; in this case we refer to it as setting the spin to equal the sign of the corresponding entry of the vector.  Alternatively, if the entries of the vector are constrained to the interval $[-1,+1]$, we can choose the final spin configuration, choosing each spin independently and choosing the $i$-th spin so that the expectation of $Z_i$ is equal to the $i$-th component of the vector.

To write this formally, let us define vectors $\vec v_a$, for $a=0,1,\ldots$.  The initial vector will be $\vec v_0$ and each $\vec v_{a+1}$ will be defined from $\vec v_a$ by the local  transformation depending on the objective function and the nonlinear transformation acting on each entry independently.

To define the local transformation, if the objective function that we seek to minimize is given by
$\frac{1}{2}\sum_{ij} J_{ij} Z_i Z_j$, then we update by applying a linear transformation
$$\vec v_a+c_a J\cdot \vec v_a,$$ for some scalar $c_a$ depending on the step number $a$.  To define the nonlinear transformation, let $g_a(\cdot)$ be some function from real numbers to real numbers, depending on the step number $a$.
We then define
\be
(\vec v_{a+1})_i=g_{a}\Bigl((\vec v_a+c_a J \cdot \vec v_a)_i\Bigr),
\ee
where the subscript $i$ after parenthesis denotes the $i$-th entry of a vector.
More generally, we can allow $g_a$ to depend upon some additional randomness, with independent randomness on each site; an example of this is given below.

Finally, after some number $b$ of steps, we have some final $\vec v_b$, and we apply a final transformation to determine a vector of spins $\vec Z$.

More generally, if the objective function is a cubic or higher degree function of the vector $\vec Z$, then we define the transformation as follows.  In the cubic degree case, such as an objective function $\frac{1}{3!}\sum_{i,j,k} J_{i,j,k} Z_i Z_j Z_k,$
then for any given vector of spins $\vec Z$, define the ``force"\footnote{The term ``force" is used because it is, roughly, a derivative of the objective function (which we can regard as an ``energy") with respect to a variable.}
$\vec F_a$ to be a vector with entry $$(\vec F_a)_i=\frac{1}{2} \sum_{j,k} J_{i,j,k} (\vec v_a)_j (\vec v_a)_k.$$  More generally, for objective functions depending on a product of $4$ or more entries of $\vec Z$, we define $(\vec F_a)_i$ to equal one-half of the difference between the value of the objective function with $Z_i=+1$ and the value with $Z_i=-1$, replacing each occurrence of $Z_j$ for $j\neq i$ in the polynomial defining the objective function with the continuous variable $(\vec v_a)_j$.
Then, we update with
\be
(\vec v_{a+1})_i=g_{a}\Bigl((\vec v_a+c_a \vec F_a)_i \Bigr),
\ee
where $\vec F_a$ is the vector of forces for the given $\vec v_a$.

This description of the classical algorithm might seem rather general and also may be difficult to understand.  Thus, before generalizing it further in section \ref{generalize}, we
will explain several specific instances of this algorithm in 
section \ref{M3L2} where we analyze the algorithm for MAX-3-LIN-2, and in section \ref{TF} where we apply it to MAX-CUT on a triangle-free graph, showing also that one instance of this algorithm is that of Ref.~\onlinecite{hirvonen2014large}, but we also use this framework to generalize.
In both of these cases, we will consider the simplest form of the algorithm with only a single step.

Before, however, giving these specific examples, let us relate this algorithm to simulated annealing, explaining that simulated annealing is an instance of this algorithm.
A simple simulated annealing algorithm\ocite{aarts1988simulated} with local spin flips proceeds by initializing the system at infinite temperature, i.e., initializing each spin independently, choosing the value of the spin from the uniform distribution on $\{-1,+1\}$.  Then, the algorithm proceeds through a series of time steps; on each time step, spins are randomly updated following Boltzmann transition rules.  That is, each spin is randomly flipped with the acceptance probability depending upon Boltzmann weights, i.e., the acceptance probability depends upon the force upon that spin.  The temperature for these weights decreases as a function of the step number.
This is an instance of a local tensor algorithm: choose entries of $\vec v_0$ independently from the uniform distribution on $\{-1,+1\}$.  Then, we will apply a transformation so that $\vec v_a$ represents the spin configuration after $a$ steps of the annealing algorithm.  To do this, choose $c_a>0$ small so that $\vec v_a + c_a \vec v_a$ has absolute value close to $1$ but so that the magnitude of the spin is slightly shifted by an amount proportional to the force on that spin.  Then apply a transformation $g_a$ incorporating randomness whose image is $\{-1,+1\}$; since the input to the random function contains both the sign of the spin after $a$ steps as well as the force on that spin, it is possible to choose this function so that the resulting probability indeed is chosen from the Boltzmann distribution.

Remark: one minor difference between this local tensor algorithm and a typical annealing algorithm is that the updates are done on all spins in parallel.  This gives a slight difference to the annealing algorithm where typically a single spin is updated at a time.  However, we can change the function $g_a$ to reduce the update probability, i.e., keeping the temperature the same but reducing the transition rate; in the limit of small update probability, this reduces to sequential updates of the spins.

The ability of this algorithm to reproduce simulated annealing means that, in principle, with sufficiently many steps, the algorithm can exactly solve (with probability arbitrarily close to $1$) any MAX-K-LIN-2 instance.  See the discussion for some comments on why this is and is not interesting.

\section{MAX-3-LIN-2 }
\label{M3L2}
We now apply such a local tensor algorithm to the case of MAX-3-LIN-2.  The algorithm will give the same scaling performance as the one-step QAOA for this problem; see Eq.~(\ref{m3l2reseq}) below.
We assume that the tensor $J_{i,j,k}$ has entries chosen from $\{-1,0,+1\}$ and that every spin $i$ participates in $D$ distinct nonzero terms.
The normalization $1/3!$ is placed in front since we choose $J_{i,j,k}$ symmetric in its indices.

Choose entries of $(\vec v_0)_i$ independently and uniformly from $\{-1/2,+1/2\}$.  Here the choice of $\pm 1/2$ is simply a convenient choice which could be optimized better to improve constant factors in the algorithm.
Then, let
$(\vec v_1)_i=(\vec v_0+c_0 \vec F_0 \Bigr)_i$ where $c_0$ is chosen below (that is, $g_0$ is simply the identity function, so no nonlinear transformation is applied).  Finally apply the following rule to determine $Z_i$: if $|(\vec v_1)_i|> 1$, then choose $Z_i$ arbitrarily (a natural choice is to choose $Z_i$ to equal the sign of $(\vec v_1)_i$ but for the proof below the choice does not matter).  Otherwise, if $|(\vec v_1)_i| \leq 1$, choose $Z_i$ randomly so that its expectation value is equal to $(\vec v_1)_i$; the different $Z_i$ are chosen independently.
This method of choosing $Z_i$ from $v_i$ for $|v_i|\leq 1$ is sometimes referred to as using ``soft spins".

Our main result in this section is that
\be
\label{m3l2reseq}
\expec[\frac{1}{3!} \sum_{i,j,k} J_{i,j,k} Z_i Z_j Z_k]=\Theta(D^{1/4} N),
\ee where
$\expec[\ldots]$ denotes expectation value. 
Since there are $DN/3$ terms in the objective function, the improvement is $\Theta(D^{-3/4})$ per term, matching the scaling of the one-step QAOA (in this section we are not concerned with optimizing the prefactor in the scaling, in contrast to the next section for triangle-free MAX-CUT where we do optimize the prefactor).

To get oriented, let us first ignore the case with $|(\vec v_1)_i|> 1$.  Let us simply imagine that it is possible to choose $Z_i$ so that its expectation value is equal to $(\vec v_1)_i$, even though this is only possible for $|(\vec v_1)_i|\leq 1$.
In this approximation, the expectation value of the objective function is equal to the expectation value with respect to the continuous variables in $\vec v_1$, i.e.,
$\frac{1}{3!}\expec[\sum_{i,j,k} J_{i,j,k} (\vec v_1)_i (\vec v_1)_j (\vec v_1)_k].$
We will evaluate this expectation value and then later we bound the probability that $|(\vec v_1)_i|>1$.
We have:
\begin{eqnarray}
\label{objexpeq}
&&\frac{1}{3!}\expec[\sum_{i,j,k} J_{i,j,k} (\vec v_1)_i (\vec v_1)_j (\vec v_1)_k] \\ \nonumber
&=& 
\frac{1}{3!}\expec[\sum_{i,j,k} J_{i,j,k} (\vec v_0)_i (\vec v_0)_j (\vec v_0)_k] \\ \nonumber
&& +
c_0\frac{1}{2}\expec[\sum_{i,j,k} J_{i,j,k} (\vec F_0)_i (\vec v_0)_j (\vec v_0)_k] \\ \nonumber
&& +c_0^2\frac{1}{2}\expec[\sum_{i,j,k} J_{i,j,k} (\vec F_0)_i (\vec F_0)_j (\vec v_0)_k] \\ \nonumber
&&+c_0^3 \frac{1}{3!}\expec[\sum_{i,j,k} J_{i,j,k} (\vec F_0)_i (\vec F_0)_j (\vec F_0)_k],
\end{eqnarray}
where the factors of $1/2=3/(3!)$ result from symmetry: we have
$\expec[\sum_{i,j,k} J_{i,j,k} (\vec F_0)_i (\vec v_0)_j (\vec v_0)_k]+\expec[\sum_{i,j,k} J_{i,j,k} (\vec v_0)_i (\vec F_0)_j (\vec v_0)_k]+
\expec[\sum_{i,j,k} J_{i,j,k} (\vec v_0)_i (\vec v_0)_j (\vec F_0)_k]
=
3\expec[\sum_{i,j,k} J_{i,j,k} (\vec F_0)_i (\vec v_0)_j (\vec v_0)_k]$.
The reader should not worry too much about these combinatorial factors such as $1/3!$.  These only affect a constant prefactor in the performance and we are not attempting to optimize that constant.

The terms 
$\frac{1}{3!}\expec[\sum_{i,j,k} J_{i,j,k} (\vec v_0)_i (\vec v_0)_j (\vec v_0)_k]$
and
$\frac{1}{2}\expec[\sum_{i,j,k} J_{i,j,k} (\vec F_0)_i (\vec F_0)_j (\vec v_0)_k]$ both vanish.  Note that $(\vec F_0)_i$ is a polynomial which is homogeneous of degree $2$ in the
entries of $\vec v_0$,
so that the latter expectation value is a polynomial which is homogeneous of degree $5$ in the entries of $\vec v_0$ so it vanishes on average.

We have
\begin{eqnarray}
\label{exp1}
c_0\frac{1}{2}\expec[\sum_{i,j,k} J_{i,j,k} (\vec F_0)_i (\vec v_0)_j (\vec v_0)_k]&=&\frac{c_0}{4}\sum_{i,j,k,l,m} J_{i,j,k} J_{i,l,m} \expec[(\vec v_0)_j (\vec v_0)_k (\vec v_0)_l (\vec v_0)_m]\\ \nonumber
&=&
\frac{c_0}{16}\sum_{i,j,k,l,m} J_{i,j,k} J_{i,l,m} (\delta_{j,l}\delta_{k,m}+\delta_{j,m}\delta_{k,l}) \\ \nonumber
&=&
\frac{c_0}{8} \sum_{i,j,k} J_{i,j,k}^2\\ \nonumber
&=& \frac{c_0}{4}DN,
\end{eqnarray}
where the last equality uses that there are $N$ choices of $i$ and for each such choice there are $2D$ choices of $j,k$ giving a nonvanishing result (there are $2D$ choices because $J_{i,j,k}$ is symmetric under interchange of $j,k$)

\begin{figure}
\includegraphics[width=2.5in]{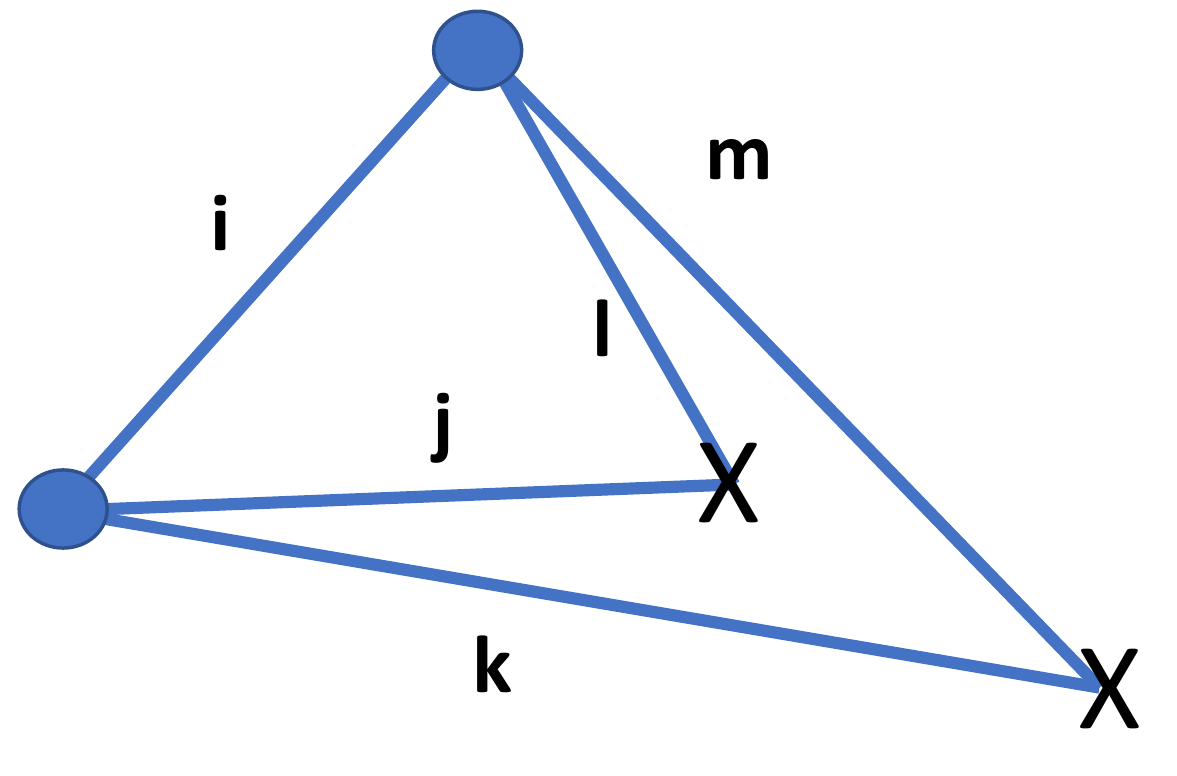}
\caption{Diagram for expectation value in Eq.~(\ref{exp1}).}
\label{figd1}
\end{figure}

These expectation values may be written using diagrams similar to those in quantum field theory,
as shown in Fig.~\ref{figd1}.  
This diagram is a tensor network whose contraction gives the desired expectation value.
This diagrammatic interpretation is not needed and may be skipped, but it gives a way to keep track of indices that may be useful for readers with a quantum field theory background.  Each solid circle represents a tensor $J$.  We have labelled indices $i,j,\ldots$ on the lines.  An ``X" in the figure denotes requiring two indices to be equal, using $\expec[(\vec v_0)_k (\vec v_0)_m]=\delta_{k,m}$.  Thus an ``X" represents a two-index tensor on which both indices must be equal (so, we can trivially remove the ``X" from the tensor network).
Note that we have 
$\expec[(\vec v_0)_j (\vec v_0)_k (\vec v_0)_l (\vec v_0)_m]=\frac{1}{4}(\delta_{j,l}\delta_{k,m}+\delta_{j,m}\delta_{k,l}+\delta_{j,k}\delta_{l,m}-2\delta_{j,k} \delta_{k,l}\delta_{l,m})$.  The last two terms vanish when multiplied by $J_{i,j,k} J_{i,l,m}$.
The figure shows just the first choice $\delta_{j,l}\delta_{k,m}$; the second choice must be added in as well.  Remark: these two different choices give the same diagram up to some relabelling of the edges so they evaluate to the same value; for more complicated diagrams this will lead to more complicated combinatorial factors.

Remark: for those interested in using such diagrams further, note that if we had instead $\expec[(\vec v_0)_j (\vec v_0)_k (\vec v_0)_l (\vec v_0)_m]=\frac{1}{4}(\delta_{j,l}\delta_{k,m}+\delta_{j,m}\delta_{k,l}+\delta_{j,k}\delta_{l,m})$ this would be the expectation value for entries of $(\vec v_0)_i$ chosen from a Gaussian distribution; the added term $\delta_{j,k} \delta_{k,l}\delta_{l,m}$ represents deviation from Gaussianity and could be written diagrammatically using an ``X" with four lines attached.

We can bound the absolute value of the last term in Eq.~\ref{objexpeq} as follows.
This is equal to 
\begin{eqnarray}
\label{4expec}
&& c_0^3\frac{1}{3!}\expec[\sum_{i,j,k} J_{i,j,k} (\vec F_0)_i (\vec F_0)_j (\vec F_0)_k] \\ \nonumber
&=&
c_0^3 \frac{1}{3!} \frac{1}{2^3}\sum_{i,j,k,l,m,n,o,p,q} J_{i,j,k} J_{i,l,m} J_{j,n,o} J_{k,p,q} \expec[(\vec v_0)_l
(\vec v_0)_m
(\vec v_0)_n
(\vec v_0)_o
(\vec v_0)_p
(\vec v_0)_q].
\end{eqnarray}
Consider the expectation value $\expec[(\vec v_0)_l
(\vec v_0)_m
(\vec v_0)_n
(\vec v_0)_o
(\vec v_0)_p
(\vec v_0)_q]$.  This vanishes if some site appears an {\it odd} number of times in the expectation value, e.g., if there is some $r$ such that $\delta_{l,r}+\delta_{m,r}+\delta_{n,r}+\delta_{o,r}+\delta_{p,r}+\delta_{q,r}$ is odd.  So, to have a non-vanishing expectation value, it must be the case that $l$ is equal to one of the other indices $m,n,o,p,q$.  However, if $l=m$, the expectation value vanishes.  One way for the expectation value to be nonvanishing is to have $m=n,o=p,q=l$.  

There are other possible ways to have a nonvanishing expectation value: these amount to choosing a perfect matching on the
six variables, $l,m,n,o,p,q$ such that the matching does not match $l$ with $m$ or $n$ with $o$ or $p$ with $q$ (since in this case, the sum vanishes).  There are $8$ such matchings, 
and for such matchings, the expectation value is equal to $(1/2)^3$.
So, 
\begin{eqnarray}
&&c_0^3 \frac{1}{3!} \frac{1}{2^3}\sum_{i,j,k,l,m,n,o,p,q} J_{i,j,k} J_{i,l,m} J_{j,n,o} J_{k,p,q} 
 \expec[(\vec v_0)_l
(\vec v_0)_m
(\vec v_0)_n
(\vec v_0)_o
(\vec v_0)_p
(\vec v_0)_q]\\ \nonumber
&=&c_0^3 \frac{1}{3!} \frac{1}{2^3} \sum_{i,j,k,l,m,n} J_{i,j,k} J_{i,l,m} J_{j,m,o} J_{k,o,l}.
\end{eqnarray}
One might worry that we have overcounted: if for example we consider the two different matchings $m=n,o=p,q=l$ and $m=p,q=n,o=l$, these might in some cases correspond to the same term in the sum over $l,m,n,o,p,q$.  However, for these two matchings to correspond to the same terms in the sum we must have $m=n=o$ and so $J_{j,n,o}$ vanishes (remark: even if we allowed nonvanishing $J_{j,n,o}$, such terms would be lower order in $D$).

\begin{figure}
\includegraphics[width=2.5in]{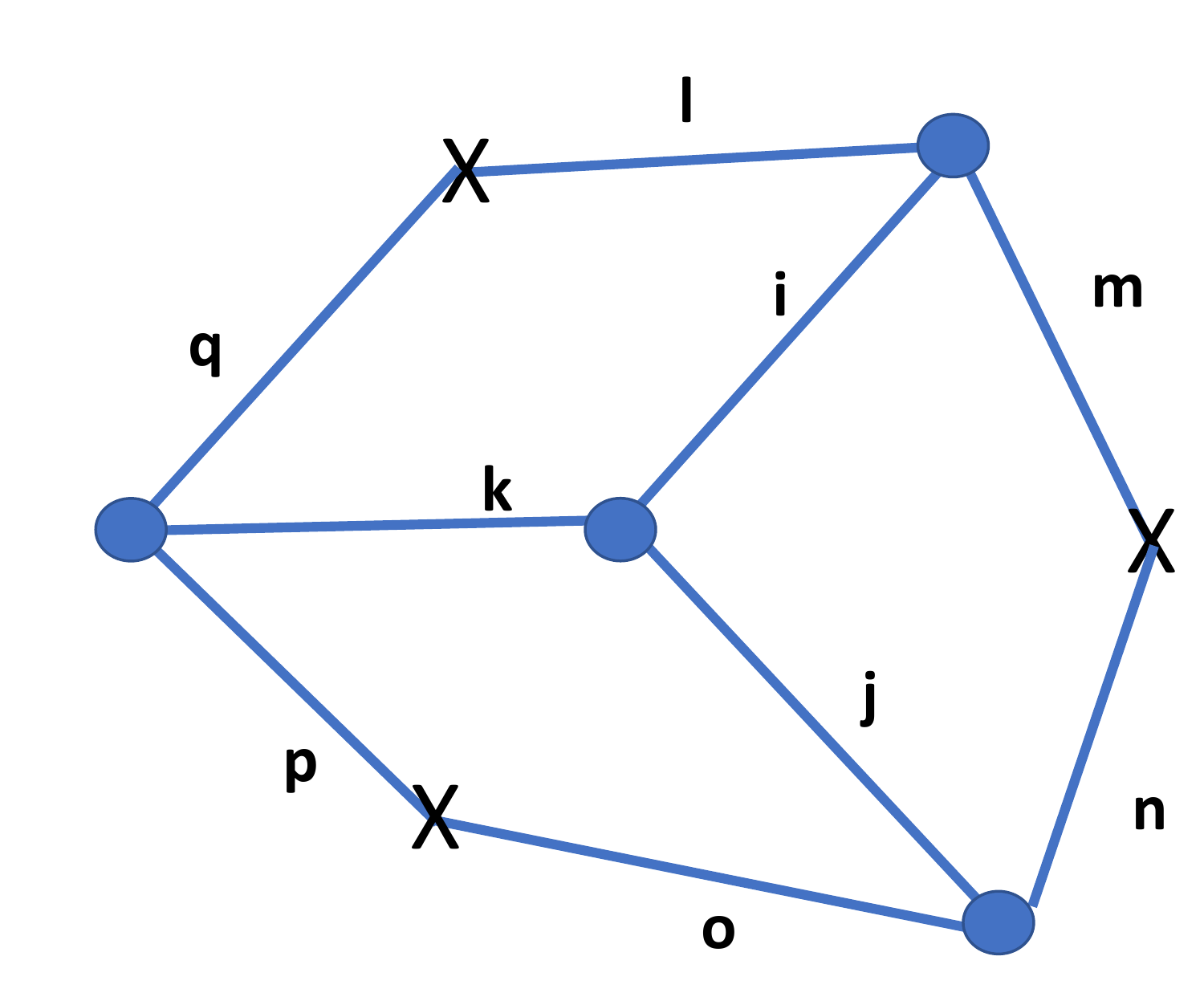}
\caption{Diagram for expectation value in Eq.~(\ref{4expec}).}
\label{figd2}
\end{figure}

This more complicated sum can also be interpreted diagrammatically as shown in Fig.~\ref{figd2}.

We now bound $\sum_{i,j,k,l,m,n} J_{i,j,k} J_{i,l,m} J_{j,m,o} J_{k,o,l}$.
There are $N$ possible choices of $i$.  For each such choice there are $2D$ choices of $j,k$ which give a nonzero result.  Hence, there are $2DN$ choices of $i,j,k$.  Given these choices, let $A_1,A_2,A_3$ be three matrices, with $A_1$ having matrix elements $(A_1)_{lm}=J_{i,l,m}$ and similarly with $(A_2)_{m,n}=J_{j,m,n}$ and $(A_3)_{n,l}=J_{k,n,l}$.
Then, $\sum_{l,m,n} J_{i,l,m} J_{j,m,n} J_{k,n,l}={\rm tr}(A_1 A_2 A_3)$.  Each matrix $A_i$ has $2D$ nonzero elements, each of which has absolute value $1$.  We claim then that the trace is bounded by $(2D)^{3/2}$.  

This bound is the consequence of a more general result for tensor networks (this trace is a tensor network):
\begin{lemma}
\label{onlylemma}
Consider any tensor network $M$ with $N_T$ tensors in the network, so that each tensor $T$ has at most $d$ nonzero entries and so that these entries are 
bounded by $1$ in absolute value.
 Then,
the contraction of the tensor network is bounded in absolute value by $d^{N_T/2}.$
\begin{proof}
The result is immediate if the graph corresponding to the tensor network is bipartite, as then one of the two parts of the graph has at most $N_T/2$ tensors and there are at most $d^{N_T/2}$ choices of labels on the edges which give a nonzero value.

We will repeatedly use the following construction to reduce to the bipartite case.
Given a tensor network $M$, and given any 
set of vertices $S$, we construct a new network $\Cut(M,S)$ as follows.  Cut all edges in $M$ which connect vertices in $S$ to vertices in the complement of $S$, giving a new network with external edges.  Join this network to a copy of itself on the external edges, joining external edges on each vertex in one copy to the corresponding edges on the other copy of that vertex.
The resulting network is $\Cut(M,S)$.

Let $\Val(\cdot)$ denote the scalar resulting from contracting some tensor network.
By Cauchy-Schwarz,
$\Val(M)\leq \sqrt{\Val(\Cut(M,S))}$.
Clearly, $|\Cut(M,S)|=2|M|$, where absolute value signs denote the numbers of vertices in a network. 
Hence, if we can construct a sequence of networks $M_1=\Cut(M,S_1)$, $M_2=\Cut(M,S_2)$, and so on, for some sets $S_i$, so that the final network in the sequence is bipartite, then the lemma follows.

One way to construct this sequence is to label the vertices in $M$ by integers $1,2,\ldots,N_T$.  Then, let $S_1=\{1\}$.  Let $S_2$ denote both copies of vertex $2$, and in general let $S_i$ denote all $2^i$ copies of vertex $i$.  Then $S_{N_T}$ is bipartite, consisting of $N_T$ disconnected hypercubes.
Remark: one can also construct a bipartite network using only 
$\ell \equiv \lceil \log_2(N_T) \rceil$ steps by labelling the vertices in $M$ by distinct bit strings of length $\ell$.
Then, on the $i$-th step, let $S_i$ denote all copies of vertices with a $1$ in the $i$-th position in the bit string.
\end{proof}
\end{lemma}

Thus, the term
$c_0^3\frac{1}{3!}\expec[\sum_{i,j,k} J_{i,j,k} (\vec F_0)_i (\vec F_0)_j (\vec F_0)_k]$ is bounded in absolute value by
$$\frac{1}{3!} \frac{1}{2^3} DN (2D)^{3/2} |c_0|^3=O(ND^{5/2}|c_0|^3).$$

Hence if we choose $c_0$ to equal some positive constant times $D^{-3/4}$, the second term on the right-hand side of Eq.~(\ref{objexpeq}) is $\Theta(D^{1/4} N)$ and the remaining terms are bounded in absolute value by $O(D^{1/4} N)$.  
Choosing this constant sufficiently small makes these remaining terms smaller compared to the second term by any desired factor, so we can choose $c_0$ so that
\be
\frac{1}{3!} \expec[\sum_{i,j,k} J_{i,j,k} (\vec v_1)_i (\vec v_1)_j (\vec v_1)_k] = \Theta(D^{1/4} N).
\ee

Up to this point we have ignored the possibility that $|(\vec v_1)_i|>1$.  We now bound the probability that this occurs for a given site $i$, hence bounding the expected number of sites for which it occurs and bounding the difference 
$\expec[\sum_{i,j,k} J_{i,j,k} (\vec v_1)_i (\vec v_1)_j (\vec v_1)_k]- \expec[\sum_{i,j,k} J_{i,j,k} (\vec v_1)_i (\vec v_1)_j (\vec v_1)_k]$.
Indeed, to have $|(\vec v_1)_i|>1$, we must have $|F_i|\geq (1/2) c_0^{-1}$.
The probability bound that we use is very similar to one used in Refs.~\onlinecite{barak2015beating,qqdq}.

We use the fact that
by theorem 9.23 of Ref.~\onlinecite{o2014analysis}, for any function $f$ of degree at most $K$ from $\{-1,1\}^N\rightarrow \mathbb{R}$ we have for any
$t \geq (2e)^{K/2}$ that
\be
\label{boundeq}
\pr_{x \in \{-1,1\}^N}[|f(x)|\geq t \expec[|f|^2]^{1/2}] \leq \exp(-\frac{K}{2e} t^{2/K}).
\ee
The force $F_i$ is a function of degree $K=2$ in this case (for more general MAX-K-LIN-2, the force $F_i$ has degree $K-1$).
We have $\expec[F_i^2]^{1/2}=\sqrt{D}$.
Hence for $c_0=O(D^{-3/4})$, we can apply the above bound with $t=\Omega(D^{1/4})$ so the probability that
$|F_i|\geq (1/2) c_0^{-1}$ is bounded by $\exp(-\Omega(D^{1/4}))$.  This exponentially decaying function of $D$ is negligible for large $D$
so the result  Eq.~(\ref{m3l2reseq}) follows.

This result generalizes straightforwardly to MAX-K-LIN-2 for odd $K>3$.
Using the same algorithm,
the expectation value of the objective function is $\Theta(D^{1/4} N)$, where we
ignore $K$-dependent prefactors.
The case in which we have $|(\vec v_1)_i|>1$ occurs with negligible probability as before.

When evaluating the expectation value
$\expec[\sum_{i_1,\ldots,i_K} J_{i_1,\ldots,i_K} (\vec v_1)_{i_1} (\vec v_1)_{i_2} \ldots (\vec v_1)_{i_K}],$
again each $(\vec v_1)_{i_a}$ is a sum of two terms, $(\vec v_0)_{i_a}$ and $(\vec F_0)_{i_a}$.
Consider a given term in the expansion where we include the force a total of $m$ times.
All terms with $m$ even vanish in expectation.
For $m=1$, the result is $\Theta(c_0DN)$ as before.
For arbitrary $m>1$, we obtain an equation similar to Eq.~(\ref{4expec}), summing over choices of $K+m(K-1)$ distinct indices.  Each term in the sum has $m+1$ factors of $J$ multiplying
an expectation value of a 
product of entries of $\vec v_0$.  This expectation value is a sum of products of $\delta$-functions.
Call any term with the minimal number of $\delta$-functions in the product a ``minimal term".  Call the terms with more $\delta$-functions ``non-minimal terms".
Any minimal term
is equal to $c_0^m$
times some tensor network with $m+1$ tensors, each tensor having $K$ legs and being equal to $J$.  
Any non-minimal term can also be written as a tensor network, but we must now introduce tensors with $4$ or more legs to denote certain products of $\delta$-functions.  However, we can bound the absolute value of the non-minimal terms as follows: without loss of generality, assume that all nonzero entries in $J$ have the same sign.  Then if we remove some number of $\delta$-functions from the product in a nonminimal term, for example replacing $\delta_{j,k} \delta_{k,l}\delta_{l,m}$
with $\delta_{j,k}\delta_{l,m}$, this does not reduce the absolute value of the expectation value.  So the absolute value of any nonminimal term is bounded by some the absolute value of some minimal term.

A minimal term corresponds to a tensor network $L$ with $m+1$ vertices.  
Fix some choice of $i_1,\ldots,i_K$ which gives a nonzero value to $J_{i_1,i_K}$.  There are $O(DN)$ such choices.
For each choice, the contraction over the other indices in $L$ gives some tensor network $L'$ with $m$ tensors.
Hence any minimal term with $m>1$ is bounded by
$O(c_0^m D N D^{m/2}),$ by lemma \ref{onlylemma} and for odd $K$ the term with $m=2$ vanishes.
So, as before we can choose $c_0$ proportional to $D^{-3/4}$ so that 
the expectation value of the objective function is $\Theta(D^{1/4} N)$.

In contrast, for even $K$, the term with $m=2$ may be present so the expectation value of the objective function is only $\Theta(N)$.

\section{MAX-CUT on a Triangle Free Graph}
\label{TF}
As another comparison, we consider MAX-CUT on a triangle-free graph of degree $D$.  We compare the QAOA with a single step to a local classical
algorithm.

We can regard an instance of MAX-CUT as MAX-2-LIN-2.  Each vertex $i$ of the graph corresponds to a spin $\sigma_i=\pm 1$, with the sign of the spin determining the cut.  The coupling matrix is $J_{ij}=0$ if $i,j$ are not connected by an edge and $J_{ij}=-1$ if they are.

Let $N_e$ be the total number of edges on the graph.
Then, given a vector of spins $\vec \sigma$, the number of edges cut by the corresponding cut is equal to $\frac{1}{2} N_e +\frac{1}{4} \vec \sigma \cdot J \cdot \vec \sigma$.
There is a factor of $1/4$ in front since $J$ is a symmetric matrix.

The optimal value of the parameters of the single step QAOA can be computed analytically, giving an expected fraction of edges cut (i.e., number of cut edges divided by $N_e$) equal to 
$$\frac{1}{2}+\frac{1}{2\sqrt{D}} \Bigl(1-\frac{1}{D}\Bigr)^{\frac{D-1}{2}} \geq \frac{1}{2}+\frac{0.3032}{\sqrt{D}},$$
as shown in Refs.~\onlinecite{wang2018quantum,ryan2018quantum}.
In general, when describing the performance of these algorithms, we will say that an algorithm improves by a factor $\delta$ over random if the expected fraction of cut edges is at least $1/2+\delta$.

It was claimed\ocite{ryan2018quantum} that this value
``improves upon the currently known best classical approximation algorithm for these graphs".
We show that this is not true.  Of course, it is likely that many algorithms (for example,  Goemans-Williamson) may beat this QAOA algorithm, but proving this may be difficult and also Goemans-Williamson is not a local algorithm.  However, we will show that a local classical algorithm beats this QAOA for {\it all} values of $D$ and further that for most choices of $D$, such a classical algorithm was already known!  The algorithm of Ref.~\onlinecite{hirvonen2014large} is such a local algorithm.  As explained in that reference, the algorithm consists of randomly assigning each spin to a cut.  Then, for each spin, if sufficiently many neighbors {\it agree} with the spin (i.e., are in the same cut), the given spin changes to which cut it is assigned.
The study of this algorithm requires optimizing a single parameter, the threshold at which the spin changes to which cut it is assigned.  This threshold is denoted $\tau$.  We will refer to this algorithm as the ``threshold algorithm" below.

Precise analytical bounds were given on the performance in Ref.~\onlinecite{hirvonen2014large} for a particular choice of this threshold.  With this choice of threshold, the algorithm asymptotically has an expected fraction of cut edges greater than or equal to 
$\frac{1}{2}+\frac{0.2812}{\sqrt{D}}$, i.e., it improves by at least $\frac{0.2812}{\sqrt{D}}$ over random (the value given in that reference was actually a lower bound to the improvement, so it is possible that the actual improvement is larger).

However, if one simply numerically optimizes the threshold, which can be done very simply as we explain below, we find that for all choices of $3\leq D\leq 1000$ that this threshold algorithm outperforms the QAOA with the exceptions of $D=3,4,6,11$ for which the QAOA is better (for $D=2$, both the algorithms have the same performance).  In appendix \ref{numres} we give some numerical results for $D$ up to $19$. Further, a numerical scaling analysis given below suggests that the classical algorithm outperforms the quantum algorithm for all $D>1000$; it seems like it would not be difficult to give a precise proof of this using asymptotic properties of a binomial distribution but we do not give this here.
Indeed, we emphasize such a numerical optimization was {\it already given} in Ref.~\onlinecite{hirvonen2014large}, although no comparison to QAOA was done since that paper predates the QAOA.

This leaves open the cases of $D=3,4,6,11$.  For these choices of $D$, we show that it is still possible to give a local classical algorithm (indeed, again with a single step) that outperforms the QAOA.  This is possible by regarding the algorithm of Ref.~\onlinecite{hirvonen2014large} as an instance of the class of local tensor algorithms considered above and then considering slightly more general choices of parameters.  For $D=6,11$ we prove that a local classical algorithm outperforms the QAOA, though we do {\it not} optimize over all such single step classical algorithms so it is likely that even further improvements are possible.  For $D=3,4$, for simplicity we just numerically sample the performance of the classical algorithm to show that it outperforms the QAOA; this ``numerical sampling" simply amounts to a Monte Carlo estimation of certain integrals which determine the exact performance.  We emphasize that here it would also likely not be difficult to prove the performance of the classical algorithm rigorously and also that again we have not optimized over possible classical algorithms so likely even further improvements are possible.  Further we give later in section \ref{numimp} a modification of the algorithm using discrete rather than continuous numbers for which we can prove that it outperforms the QAOA.

Let us first explain how to regard this classical algorithm as an instance of the local tensor algorithms above.

We choose the initial vector $\vec v_0$ with each entry chosen independently from the uniform distribution on $\{-1,+1\}$.  Then, we apply a single linear transformation with some chosen constant $c_0$ (and the transformation $g_0$ is chosen to be the identity so that $\vec v_1=\vec v_0+c_0 J \cdot \vec v_0$.  Finally, to determine the final spin configuration we choose $Z_i$ to equal the sign of $(\vec v_a)_i$.
Note then that if the $i$-th spin has $m$ neighbors which agree in the initial configuration (i.e, there are $m$ neighbors $j$ which agree, i.e., such that $(\vec v_0)_j=(\vec v_0)_i$, and $D-m$ neighbors which disagree, we have $$(\vec v_1)_i=(\vec v_0)_i \cdot \Bigl( 1-c_0(2m-D)\Bigr).$$
So, for $c_0>0$, the sign of $(\vec v_1)_i$ will differ from that of $(\vec v_0)_i$ if $m$ is large enough, i.e., if $c_0 (2m-D)>1$.  Thus, under the assumption $c_0>0$ and that for no integer choice of $m$ do we have $1+c_0(2m-D)=0$, then this local tensor algorithm is equivalent to the threshold algorithm, where $\tau$ is the small $m$ such that $c_0(2m-D)>1$.

\subsection{$D=6,11$ Algorithm}
Now, consider the cases $D=6,11$.  In both these cases a slight generalization of the threshold algorithm suffices to outperform the quantum algorithm.  Rather than taking a ``hard" threshold, so that given $m$ neighbors which agree, if $m \geq \tau$ the spin flips and if $m<\tau$ the spin does not flip, we can instead take a ``soft threshold", where the spin has some probability of flipping depending on the number of neighbors which agree.  Indeed, it sufficed to take a very simple form of this soft threshold. For $D=11$, we found that the following rule suffices.  If at most $6$ neighbors agree, then the spin does not flip.  If more than $8$ neighbors agree, then the spin does flip.  If $7$ neighbors agree, then the spin flips with a probability chosen so that its expectation value is equal to $-0.1$ times its initial value (i.e., it flips with a probability $0.55$).  This led to an improvement over random by $0.09868\ldots$, which is larger than the value $0.0936$ for the QAOA (see table in appendix \ref{numres}).  For $D=6$, we chose a similar rule.  If at most $4$ neighbors agree the spin does not flip, if $5$ or more neighbors agree the spin does flip, and if $4$ neighbors agree then the spin flips with probability $0.65$.  This led to an improvement by a factor $0.13018\ldots$, again larger than the value $0.1294$ for the QAOA.
These rules were found by a numerical search; a finer search may lead to further improvement.

Note that this is again an instance of a local tensor algorithm.  Apply the same construction as above where we showed that the threshold algorithm is an instance of a local tensor algorithm, but choose $c_0$ so that for $7$ neighbors agreeing (in the case $D=11$; the case $D=6$ is similar) we have $1+c_0(2m-D)=0$.  Then, add weak biased random noise and finally choose $Z_i$ to equal the sign of $(\vec v_1)_i$.  We choose the noise weak so that for $m\neq 7$ the noise has no effect on the sign and we choose it biased so that for $m=7$ it leads to the desired final output probabilities.

\subsection{$D=3,4$ Algorithm}
Finally, consider the cases $D=3,4$.  Here, we used the local tensor algorithm with a single step.  We chose the entries of $(\vec v_0)$ independently and uniformly from the interval $[-1,+1]$.
We then applied a single linear transformation $\vec v_1=\vec v_0+c J \cdot \vec v_0$, and set $Z_i$ to equal the sign of $(\vec v_1)_i$.
This algorithm is not an instance of the threshold algorithm: the ``threshold" needed to make the sign of $(\vec v_1)_i$ differ from that of $(\vec v_0)_i$ depends upon the magnitude of $(\vec v_0)_i$.  If the initial magnitude is larger, it is less likely for the sign to flip.  At the same time, if the initial magnitude is larger, the spin $i$ has a larger effect on the final sign of its neighbors.
We found that by choosing $c=0.6$ for $D=3$ we obtained an improvement over random $0.1980\ldots$ and for $D=4$ by choosing $c=0.54$ we obtained an improvement over random by $0.1693\ldots$, in both cases greater than QAOA which has improvement over random
$0.1925$ for $D=3$ and
$0.1624$ for $D=4$.

It is interesting to speculate {\it why} this uniform distribution works better than the discrete distribution $\{-1,+1\}$ that gives a threshold algorithm.  One possibility may be the way in which the magnitude of a spin can determine both how easily it is flipped and its influence on its neighbors.  Indeed, consider the case $D=1$.  The simplest case of this is two spins, labelled $1,2$.  We would like them to be perfectly anti-correlated.  A single step of the QAOA can achieve this by creating an EPR pair.  However, one can also achieve it with a local tensor algorithm, if the entries of $\vec v_0$ are chosen from $[-1,1]$ or other uniform distribution, and we use constant $c_0=1$.  Then, we have $(v_1)_1=-(v_1)_2$ and whichever spin has the larger initial magnitude will not flip while its neighbor will become anti-correlated with it.

Another reason why this uniform distribution works well might simply be that now $c$ is a continuous parameter that can be tuned to optimize the algorithm.  For the threshold algorithm, there is only a single discrete parameter $\tau$ that can be adjusted and perhaps, ``by accident", for certain small choices of $D$ it is hard to adjust this parameter well.  Indeed, some of the numerical data in Fig.~\ref{figperf} later show oscillations in the performance of the threshold algorithm as a function of $D$ and perhaps for small sizes one simply is ``unlucky" with respect to some of these oscillations.

We note that we did {\it not} do an extensive search over different choices of distributions.  The uniform distribution is the first one we tried, so it is quite possible that other distributions may lead to improved performance.

\subsection{Numerical Implementation}
\label{numimp}
Now we give some implementation details to explain how the above results were obtained.  All these algorithms (QAOA and local tensor algorithms with a single step) on triangle-free graphs can be analyzed very simply, as recognized by many authors.  To compute the expectation value of $Z_i Z_j$, note that the probability distribution (or reduced density matrix in the quantum case) on any pair of neighboring spins $i,j$ can be computed given the initial distribution of the neighbors of those spins (this uses the fact that we consider only a single step of the algorithm).

Since the graph is triangle-free, the subgraph consisting of neighboring spins $i,j$ and their neighbors is the same in all cases: it is a graph with $2D$ vertices with $i,j$ connected by an edge and each connected to $D-1$ other vertices.
Thus, one simply needs to compute the expectation value of $Z_i Z_j$ given a random initial assignment $\vec v_0$ to the spins on that subgraph.

For all cases considered above, except the cases $D=3,4$, the initial random assignment is from a discrete distribution, so there are a finite number of cases to consider.  These cases can be enumerated on computer.  Of course, to speed up the enumeration, one notes that
all initial configurations related by permuting neighbors of $i$ (or related by permuting neighbors of $j$) have the same expectation value for $Z_i Z_j$.  That is, the expectation only depends on the sum of $(\vec v_0)_k$ over $k\neq j$ which are neighbors of $i$, as well as the same sum over neighbors of $j$ and also of course $(\vec v_0)_i$ and $(\vec v_0)_j$.
For the discrete distribution $\{-1,+1\}$, the probability distribution of $(\vec v_0)_k$ is readily computed from a binomial distribution.

We now give the expectation value of $\frac{1}{4} \vec \sigma \cdot J \cdot \vec \sigma$ explicitly in this case.  Let us consider a general case in which, for a given spin, if $j$ neighbors agree then the value of the spin is chosen from a distribution on $\{-1,+1\}$ with expectation value equal to $q(j) (\vec v_0)_j$, with $q(j)\in [-1,+1]$.  The threshold algorithm has $q(j) \in \{-1,+1\}$ for all $j$ while in the more general algorithm above for $D=6,11$ we choose $q(j)$ more generally.  Then,
there are two cases we consider: either $(\vec v_0)_i=(\vec v_0)_j$ or $(\vec v_0)_i=-(\vec v_0)_j$.  Each case occurs with probability $1/2$.
In the first case, the expectation value of $Z_i$ is equal to
$\sum_{n=0}^{D-1}  2^{-(D-1)} {D-1 \choose n} q(n+1) (\vec v_0)_i,$ and similarly for $Z_j$ so the expectation value of
$Z_i Z_j$ is equal to
$$\Bigl(\sum_{n=0}^{D-1}  2^{-(D-1)} {D-1 \choose n} q(n+1)\Bigr)^2,$$
where $n$ represents the number of neighbors of $i$, other than $j$, which agree with $i$.  In the second case, the expectation value of $Z_i Z_j$ is equal to 
$$-\Bigl(\sum_{n=0}^{D-1}  2^{-(D-1)} {D-1 \choose n} q(n)\Bigr)^2.$$
So,
averaging over choices of $(\vec v_0)_i, (\vec v_0)_j$, the expectation value of 
$-(1/2) Z_i Z_j$
equals
$$\frac{1}{4} \Bigl[
\Bigl(\sum_{n=0}^{D-1}  2^{-(D-1)} {D-1 \choose n} q(n)\Bigr)^2
-
\Bigl(\sum_{n=0}^{D-1}  2^{-(D-1)} {D-1 \choose n} q(n+1)\Bigr)^2
\Bigr].$$

For the cases $D=3,4$, the initial distribution is a continuous distribution.  In this case, the evaluation can be done by computing integrals.  While this evaluation likely can be done exactly, for this study we simply used a Monte Carlo method.  We randomly sampled $10^8$ configurations for each case $D=3,4$ and averaged the performance of the algorithm.  This Monte Carlo sampling leads to a statistical error of order $10^{-4}$.  The performance difference between this algorithm and the QAOA is more than $50$ times larger than this sampling error for both $D=3,4$ so that we may be very confident that indeed the algorithm outperforms the QAOA.  Also, we can approximate the continuous distribution by a discrete distribution; a simple approximation such as choosing uniformly from $\{-1,-1/3,+1/3,+1\}$ gives an algorithm whose performance can be exactly enumerated and one finds (with $c=0.599$) that it outperforms that QAOA, albeit not by as much as the continuous distribution.

\subsection{Scaling}
\begin{figure}
\includegraphics{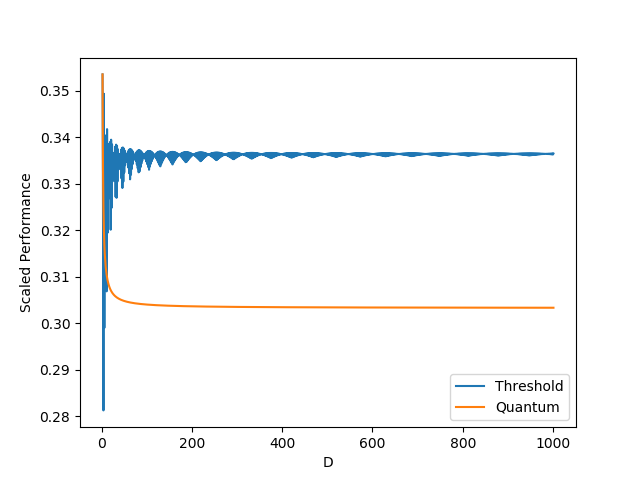}
\caption{Scaled performance of threshold algorithm and one-step QAOA.  Horizontal axis denotes degree of graph, while vertical performance indicates $\sqrt{D}$ multiplied by the improvement over a random assignment.  Note that both curves tend towards a limit as $D\rightarrow \infty$, with the threshold algorithm having a larger limit.  The threshold curve shows more ``wiggles" as a function of $D$; this may be due to the fact that one optimizes a discrete threshold and changing $D$ by one can cause the optimal threshold to jump.}
\label{figperf}
\end{figure}

We have only numerically tested the performance of the threshold algorithm up to $D=1000$, repeating the same numerical test done previously.  However, the numerical results clearly show a scaling behavior: the improvement over random approaches $c/\sqrt{D}$ for some constant $c$.  See Fig.~\ref{figperf} where we plot the improvement over random for both the threshold algorithm and QAOA.

\begin{figure}
\includegraphics{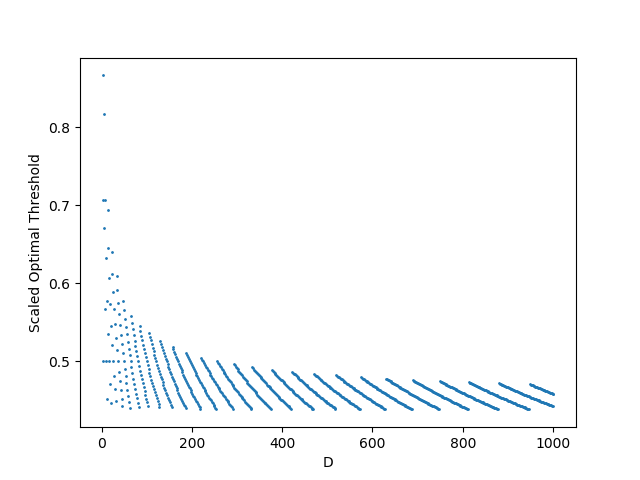}
\caption{Scaled optimal choice of threshold.  Horizontal axis denotes degree of graph, while vertical axis denotes optimal threshold minus $D/2$, divided by $\sqrt{D}$.
While the plot might not seem to represent a function (it seems like each choice of horizontal coordinate has two different vertical coordinates), this is a just a result of an even-odd oscillation in the optimal threshold which is not clearly visibly on the graph at this scale: if one plots only even $D$ or only odd $D$, one can see that indeed this represents a function.}
\label{figthresh}
\end{figure}

The optimal choice of threshold is shown in Fig.~\ref{figthresh}.  Here we see that the threshold approaches a limit $D/2+c'\sqrt{D}$ for some other constant $c'$.  It is clear from this choice of threshold why the algorithm should improve over random by an amount proportional to $c/\sqrt{D}$.  Consider a pair of neighboring spins $i,j$.  The sum $\sum_k (\vec v_0)_k$ over $k\neq j$ that neighbor $i$ can be approximated, for large $D$, by a Gaussian of width proportional to $\sqrt{D-1}$.  For the given choice of threshold, the probability that the $i$-th spin does not change its spin approaches some quantity of order unity, i.e., it limits to some nonzero number  At the same time, the probability that the sum is large enough that the $j$-th spin can ``cast the deciding vote" is of order $1/\sqrt{D}$, i.e., here we consider the probability that the sum is large enough that changing the sign of $(\vec v_0)_j$ will change the final sign of $Z_i$.  Hence, the probability that, for example, the sign of $Z_i$ remains unchanged while the sign of $Z_j$ becomes opposite to the sign of $(\vec v_0)_i$ is the product of these two probabilities and is also $\Theta(1/\sqrt{D})$.

\section{Generalizations}
\label{generalize}
We can generalize the local classical algorithm above as follows.  We emphasize that the generalizations discussed here are speculative and have not been applied to any problem.  However, this generalization may be interesting, in particular for reasons discussed later as giving another way to that a sufficient generalization of the algorithm can exactly solve any MAX-K-LIN-2; again,
see the discussion for some comments on why this is and is not interesting.

The first generalization is to consider multiple random vectors at each step.  Rather than having a sequence of random vectors $\vec v_a$, with $\vec v_{a+1}$ determined from $\vec v_a$, we can allow several random vectors $\vec v_{j,a}$ where $a$ labels that step and $j$ ranges over some discrete set of values, $1,\ldots,n_j$.  Then, the set of $\vec v_{j,a+1}$ is  determined from $\vec v_{j,a}$.  We initial each entry of $\vec v_{j,0}$, for each $j$, independently.

This is done as follows.  For MAX-K-LIN-2, $J$ is a $K$-index tensor.  For any choice of $K-1$ vectors $\vec v_{j,a}$, all having the same index $a$, we can contract $J$ with those vectors to produce a new vector.  If the index $j$ ranges over $n_j$ possible values, there are $n_j^{K-1}$ possible ways of doing this contraction.  
We then define vectors $\vec w_{j,a}$, for $1\leq j \leq n_j$, to be some linear combination of the $n_j$ different vectors $v_{j,a}$ and the $n_j^{K-1}$ vectors defined by contraction with $J$.  These linear combination is defined by an $n_j$-by-$n_j+n_j^{K-1}$ matrix.  
Then, to define $\vec v_{j,a+1}$, for each $i$ we define some function $g_a(\cdot)$ from ${\mathbb R}^{n_j}\rightarrow {\mathbb R}^{n_j}$ for each $a$. 
For each $i,a$, consider the $n_j$ different real numbers
$(\vec w_{j,a+1})_i$ given by different choices of $j$ as being entries of some vector that we denote $\vec x_{a,i}$ with $n_j$ components.
Then, set $(\vec v_{j,a+1})_i$ to be the $j$-th entry of the image of this vector under $g(\cdot)$, i.e.,
$$(\vec v_{j,a+1})_i=g_a(\vec x_{a,i})_j.$$

In the special case that $n_j=1$, this algorithm reduces to the previous algorithm.  In this case, we can drop the index $j$.  The vector that we are calling $\vec w_{a}$ then is equal to $\vec v_a+c_a F_a$.  Here we allow a slight additional generalization in that $\vec w_a$ may be an arbitrary linear combination of $\vec v_a$ and $\vec F_a$ but
a change in the dependence  on $\vec v_a$ can be absorbed into a change in $g_a$ and $c_a$ so long as the coefficient of $\vec v_a$ is nonzero.

This can be generalized even further.  Given a tensor such a $J$, we can define additional tensors by contracting tensor networks.  That is, we define some tensor network with some number of external legs, and we insert $J$ into this network; contracting the network defines a new tensor.
Choose some set of such tensor networks.
The new tensors will have some number of legs; we can then contract all but one of the legs with vectors $\vec v_{j,a}$ to give a vector, and then allow $\vec w_{j,a}$ to be an arbitrary linear combination of the resulting vectors and of the $\vec v_{j,a}$.  The algorithm above is the case that we consider only one tensor network, consisting of just tensor $J$ appearing once.

With a sufficient number of such tensor networks, even a single step can solve any MAX-K-LIN-2 to arbitrary accuracy.  To see this, imagine considering the statistical mechanics of some system with a Hamiltonian given by the negative of the objective function plus $\sum_i Z_i h_i$ for some small random $h_i$.  Suppose $h_i$ is small enough that the minimum energy solution at nonzero $h_i$ is still a minimum energy solution at $h_i=0$; the random choice of $h_i$ will however choose a unique minimum generically.
At low temperatures, the expectation value of $Z_i$ 
then converges to $\pm 1$ and it can be computed by some sum of diagrams, i.e., of tensor networks, using a series expansion in inverse temperature (see any textbook on statistical mechanics).

\section{Limitations}
We now discuss some limitations on local algorithms.  To begin, note that the one-step QAOA and the algorithm of section \ref{M3L2} both attain the same asymptotic performance on MAX-3-LIN-2 and that this is for essentially the same reasons: one adjusts some parameter (an angle in QAOA or the quantity $c_0$ in the local tensor algorithm).  At small values of this performance, the improvement over random is linear in the parameter but there is a term in the improvement over random that is cubic in the parameter and whose sign cannot be controlled.  This additional term limits one to small values of the parameter.

One might then guess that by going to additional steps in the QAOA or in the local tensor algorithm one might be able to improve the asymptotic scaling to a larger power of $D$.  In this section, we present strong evidence that this is not possible for the QAOA or for a simple version of the local tensor algorithm (some of the generalizations of the local tensor algorithms considered may evade this difficulty and may allow greater asymptotic performance with a local algorithm; we leave this as an open question).  While the result here are {\it not} a proof, they use standard semi-classical and diagrammatic arguments from physics.

We will construct an objective function so that, within a certain perturbative treatment, if the angles in the QAOA are sufficiently large, the most important contribution to the expectation value of the objective function has an uncontrolled sign.  Here, ``uncontrolled" means that we can change the sign arbitrarily by changing the sign of some  terms in the objective function.  Hence, (at least in this perturbative treatment) no guarantee can be given for the performance of the QAOA since the sign of the expectation value of the objective function is arbitrary.
On the other hand, if the angles are kept small in the QAOA, only a small improvement over random, by an amount $O(D^{1/4} N)$, is possible.

We will consider a class of objective functions that can be written in the following form:
\be
\label{Heff}
H=J_0 \ze_0^3+J_{0,1} \ze_0 \ze_1^2 + J_{1,2} \ze_1 \ze_2^2 + \ldots + J_{l-2,l-1} \ze_{l-2} \ze_{l-1}^2 +J_{l-1,l} \sum_a \ze_{l-1} \ze_{a,l}^2+J_{l,l+1}\sum_{a,p} \ze_{a,l} \ze_{a,p,l+1}^2+{\rm lower \, order \, terms}.
\ee
Some explanation of this notation is required.
Here, $J_0,J_{0,1},\ldots$ are scalars.  The quantity $l$ is an integer that we will choose later.  The choice of $l$ will depend on the number of steps in the algorithm.  The indices $a,p$ are discrete indices ranging over values $1,2,\ldots,N_a$ and $1,2,\ldots,N_p$ respectively, for some $N_a,N_p$ that we use as parameters in a perturbation theory.  However, the variables $\ze_j$ and $\ze_{a,l},\ze_{l+1}$ are not individual spins.  Rather, each is the {\it sum} of many different variables $Z_i$.  That is, for each variable $\ze_i,\ze_{a,l},\ze_{a,l+1}$ we there is some set of spins and each variable $\ze_i,\ze_{a,l},\ze_{a,p,l+1}$ is the sum of $Z_i$ over spins in this set.
Let us denote this set $S_i$ or $S_{a,l}$ or $S_{a,p,l+1}$ so that $\ze_i = \sum_{j\in S_i} Z_j$.
These sets of spins are disjoint from each other; to obtain an instance of MAX-3-LIN-2 with degree $D$, we choose each such set to have size $\Theta(\sqrt{D})$.  
The ``lower order terms" refer to the fact that when we write $\ze_2^2$, for example, this will include $\sum_{j\in S_i} Z_j^2=|S_i|$ where $|\ldots|$ denotes the cardinality of a set.  This means that we have a problem that is not MAX-3-LIN-2 since it includes terms {\it linear} in the variables.  So, we cancel these terms by adding on the same term with opposite sign; the added terms are denoted ``lower order terms" since the number of such terms is lower order in $D$.

We will choose $J_{i-1,i}=J$ for all $i$ for some scalar $J>0$ and we choose $J_0=\pm J$. 
We also choose $N_a,N_p>>1$.  The perturbation theory will controlled by parameters $N_a^{-1},N_p^{-1},D^{-1}$, all of which will be treated as small parameters in the perturbation theory; we will choose the magnitude of these perturbative quantities appropriately relative to each other as discussed below.

The perturbation theory will be constructed by expanding in $J\Phi$,
where $\Phi$ is a typical magnitude for an angle in the QAOA.  
We work in a regime with $J\Phi>>D^{-3/4}$.  We use this choice to turn an expansion in $J\Phi$ into an expansion in $D$.  Later we will also require some upper bounds on $J \Phi$; these upper bounds are to control the perturbation theory.  Of course, one might choose angles in the QAOA which are not in this regime, causing the perturbation theory to break down; however, this would require much larger angles and presently there is no evidence for that being useful.

Let us explain this angle $\Phi$ further.
The $j$-step QAOA acts on the state with all spins in the $|+\rangle$ direction with a sequence of rotations $\exp(i \theta_1 X) \exp(i \phi_1 H) \exp(i \theta_2 X) \exp(i \phi_2 H) \ldots \exp(i \theta_j X) \exp(i \phi_j H)$ to produce some state that is hoped to have a large expectation value for $H$.  Hence, we can compute the expectation value for any term in $H$ by conjugating it by these rotations.
The one-step QAOA for MAX-3-LIN-2 involves choosing $\phi=O(D^{-3/4})$.  Here, we will imagine some larger number of steps with possibly different $\phi_i$ at each step but for simplicity we imagine that all angles $\phi_i$ are roughly of the same order of magnitude: we denote the magnitude by $\Phi$.  More generally, if some of the angles $\phi_j$ are much smaller than others, we may instead decrease $l$ so that $l$ is not as large as $j+1$ but rather only equal to the number of steps with large angle, plus one.
The angles $\theta$ may be arbitrary.

We allow the scalar $J$ to be continuous, with $|J|=O(1)$.  This might seem to deviate from MAX-3-LIN-2 since in MAX-3-LIN-2, each term in the objective function is chosen with sign $\pm 1$.  If all terms are chosen with the same sign, then the resulting term such as $J_{0,1} \ze_0 \ze_1^2$ may be of order $D^{3/2}$ if $\ze_0 \sim \ze_1 \sim D^{1/2}$.
However, given a  $J_{0,1}$ with smaller magnitude, we use the following rule: for each term $Z_i Z_j Z_k$ with $Z_i \in S_0$ and $Z_j,Z_k \in S_1$ we choose the sign of the term independently from a biased distribution.  Choosing all signs the same gives the maximum magnitude, while allowing the signs to vary should (to good approximation) allow us to consider the Hamilton of Eq.~(\ref{Heff}) as a good effective approximation with smaller values of $J$.
Already, one should be surprised if a quantum algorithm can offer an advantage for such an objective function, because as $D$ becomes large, a semi-classical approximation becomes more accurate, at least for small values of the angles $\theta_i,\phi_i$.

Roughly, our idea is as follows: the terms $\sum_a \ze_{a,l} \ze_{a,p,l+1}^2$ will lead to some change in the expectation value of $\ze_{a,l}^2$ after the first step of the QAOA is applied, i.e., after we rotate by $\exp(i \theta_j X) \exp(i \phi_j H)$.  After a second step, this will lead to some change in the expectation value of $\ze_{l-1}$ due to terms
$\sum_a \ze_{l-1} \ze_{a,l}^2$.  After a third step, this will lead to a change in the expectation value of $\ze_{l-2}$ due to terms $\ze_{l-2} \ze_{l-1}^2$; after four steps this gives a change in the expectation value of $\ze_{l-3}$ due to terms $\ze_{l-3} \ze_{l-2}^2$, and so on, so after $l+1$ steps, this gives a change in the expectation value of $\ze_0$.
So, we choose $l=j-1$.
Then, this gives a contribution to the objective function proportional to this expectation value of $\ze_0^3$, multiplied by $J_0$.  Since, however, in this sequence of events, $J_0$ itself has not played any role in determining this expectation value of $\ze_0$, this will give a contribution to the expectation value of $H$ whose sign cannot be controlled by this $j$-step QAOA.

Indeed, much of this procedure can be understood semi-classically and choosing $N_a,N_p>>1$ helps justify a semi-classical approximation.  Suppose that we wish to compute $\langle \psi | \exp(-i \phi J_{i,i+1} \ze_i \ze_{i+1}^2) \exp(-i \theta X)  \ze_i \exp(i \theta X) \exp(i \phi J_{i,i+1} \ze_i \ze_{i+1}^2) | \psi \rangle$ on some state (here the state is that obtained after some number of steps of the QAOA and the angles $\phi,\theta$ are the angles used in some given step.  Then, we have $\exp(-i \theta X)  \ze_i \exp(i \theta X)=\cos(\theta) \ze_i + \sin(\theta) \yi_i$ where $\yi_i=\sum_{j\in S_i} Y_j$, and so
\begin{eqnarray}
&& \exp(-i \phi J_{i,i+1} \ze_i \ze_{i+1}^2) \exp(-i \theta X)  \ze_i \exp(i \theta X) \exp(i \phi) J_{i,i+1} \ze_i \ze_{i+1}^2)\\ \nonumber &=&\cos(\theta) \ze_i +\sin(\theta) 
\Bigl( \sin(\phi J_{i,i+1} \ze_{i+1}^2) \ex_i+
\cos(\phi J_{i,i+1} \ze_{i+1}^2) \yi_i
\Bigr),
\end{eqnarray}
where $\ex_i=\sum_{j\in S_i} X_j$.
If $\phi J_{i,i+1} \ze_{i+1}^2$ is small enough and $\ex_i$ is large, we can approximate $\sin(\phi J_{i,i+1} \ze_{i+1}^2)\approx \phi J_{i,i+1} \ze_{i+1}^2$.
Indeed, then this is a purely semi-classical update: the change in expectation value of $\ze_i$ is determined by the ``force" applied to the spin, the same as in the simplest tensor algorithm. 

For expectation value of $\ex_i$ of order $D$, this update will give a $\ze_i$ that is much larger than $\ze_{i+1}$ if
$D \phi J_{i,i+1} \ze_{i+1}^2 >> \ze_{i+1}$.  At the same time, we may still have 
$\phi J_{i,i+1} \ze_{i+1}^2<<1$ to justify the semi-classical approximation, having both of these inequalities hold so long as $\ze_{i+1}<<D$.

Here, we are approximating that the expectation value of $\ze_{i+1}^2$ is equal to the square of the expectation value.  This is of course not exactly correct.  First, even in a state that is perfectly polarized in the $X$-direction, we have that the expectation value of $\ze_{i+1}^2$ is equal to $|S_{i+1}|$, and not equal to $0$.  However, we have subtracted off the ``lower order terms" in Eq.~(\ref{Heff}) which cancels such a term proportional to $|S_{i+1}|$, leaving us with $0$.  However, it is possible to obtained a nontrivial (i.e., different from $|S_{i+1}|$) expectation value of $\ze_{a,l}^2$ due to the term $\ze_{a,l} \ze_{a,l+1}^2$ and due to a nonvanishing fourth moment of $\ze_{a,l+1}$.
Once such a nontrivial expectation value of $\ze_{a,l}^2$ is obtained, this produces an expectation value for $\ze_{a,l+1}$ which then can grow exponentially at each step.

\begin{figure}
\includegraphics[width=2.5in]{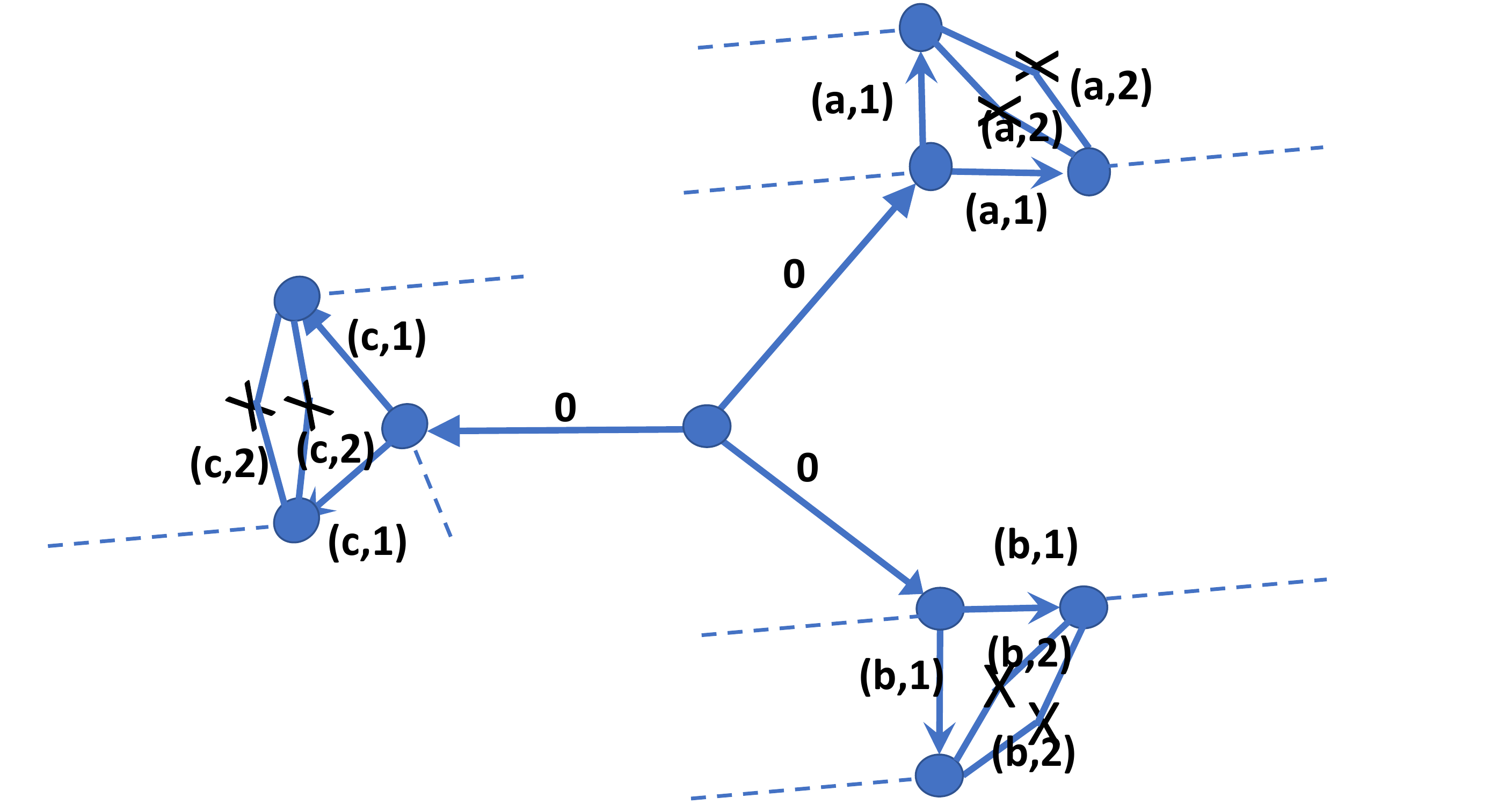}
\caption{Dominant diagrammatic contribution.  In all figures, we suppress the second index $p$ on $\ze_{a,p,l+1}$ to avoid cluttering the figure.}
\label{figd3}
\end{figure}

We can understand this diagrammatically.  Here, ``diagrammatically" simply means that we do a series expansion of the expectation value of $H$ in the angles $\phi_j$.  The objective function $H$ is a sum of products of Pauli operators.  Such a sum of products by $\exp(i \theta X)$ gives another sum of products of Paulis.  Conjugating this by $\exp(i \phi H)$ gives an infinite sum of products of Paulis when expanded in $H$.  A diagram gives a useful way to denote such sums of products.
See the diagram of Fig.~\ref{figd3}.
The lines represent different variables $\ze,\yi$; they are labelled $0,\ldots$ to indicate which variable appears on a given line.  Each circle indicates commuting with a term in $H$.  
We write an arrow on the line to indicate the order of steps: we commute a variable with the term in $H$ that the arrow point toward.
After commuting a variable $\yi$ with $H$, one may be left with a variable $\ex$.  These variable $\ex$ have an expectation $\Theta(D^{1/2})$ in the $|+\rangle$ state.  They are written as a dashed line; we do not write the label on these arrows since they may be inferred from the label on the incoming arrow.

Thus, the value of a diagram has factors depending on $\cos(\phi_j),\sin(\phi_j)$.  It has factors of $J$ depending on the circles, and a factor of $\Theta(D^{1/2})$ for each dashed line.
Finally we have drawn some ``X"s in the diagram; these indicate a pair of variables $Z_j$ (in this case, for $j\in S_{a,2})$ that coincide so that their square is equal to $1$.
Then, the diagram represents the sequence of events that we have described, reading the diagram from outside to inside: a set of four ``X"s represents a fourth moment, giving a nontrivial second moment to $\ze_{a,1}$, which gives a nontrivial first moment to $\ze_0$.  Summing over variables inside $S_{a,2}$ we see that the number of choices is also $\Theta(D^{1/2})$, so that every line with an ``X" on it gets a factor of $\Theta(D^{1/2})$.

So long as the angles $\phi_j$ are larger than $D^{-3/4}$, we can choose $J=O(1)$ so that such a diagram gives a contribution that is much larger than any linear term in $\Phi$ as follows.  Suppose we remove the dashed lines from the diagram to obtain some graph, also removing the ``X"s from the lines; this graph is a trivalent graph with no external edges so that if they have $N_v$ vertices, then they have $(3/2) N_e$ edges.  
Counting the factors of $\Theta(D^{1/2})$ for the ``X" and for the dashed lines that we have removed, we get a factor of $\Theta(D^{N_e/2})=\Theta(D^{(3/4) N_v})$.
Hence, the diagram diverges with $D$ in this regime, i.e., for $\Phi>>D^{-3/4}$, the linear term is lower order in $D^{-1}$.
We also have factors of order $\Phi^{N_v}$.

\begin{figure}
\includegraphics[width=2.5in]{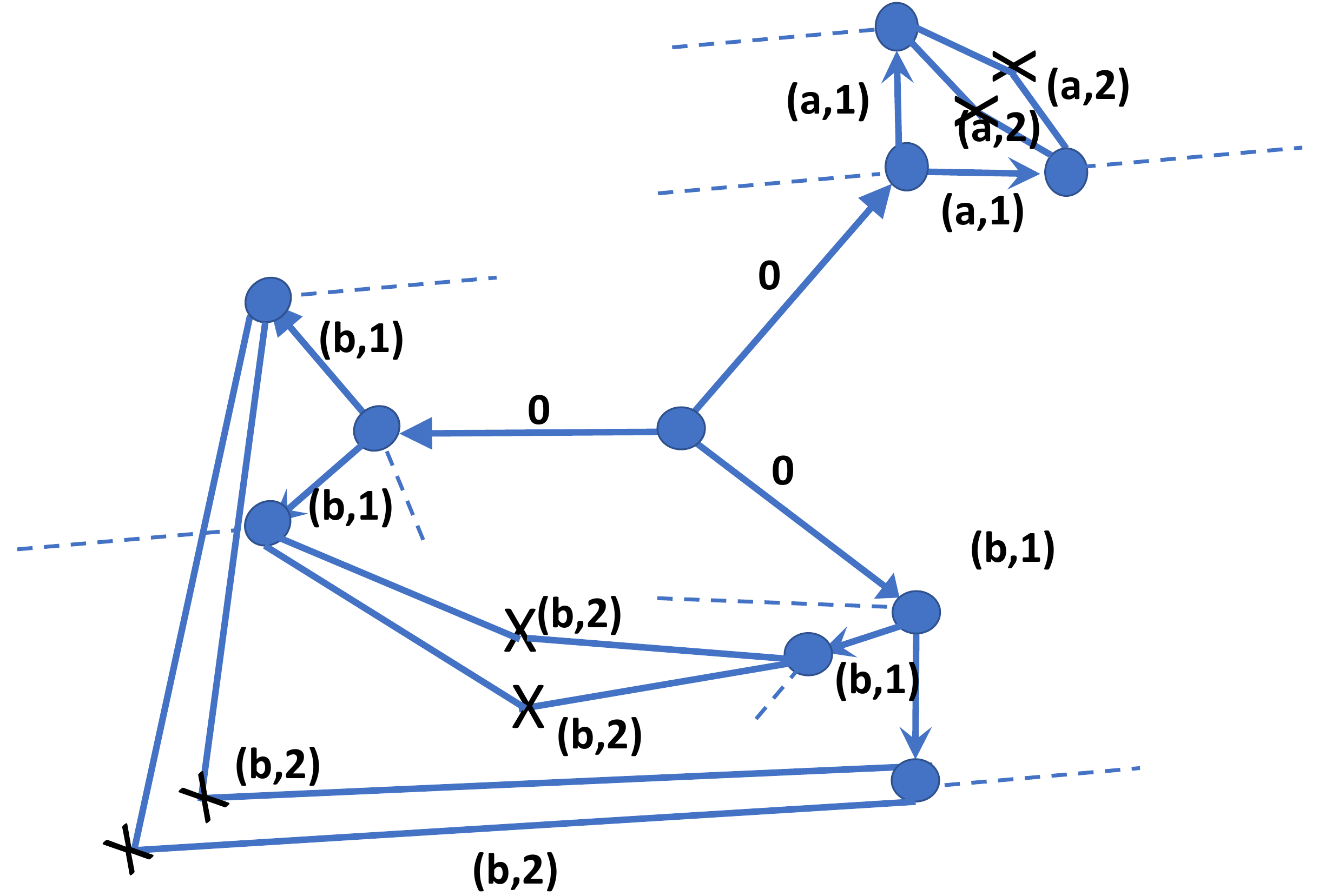}
\caption{Subleading diagrammatic contribution: lower order in $N_a$.}
\label{figd4}
\end{figure}

There are of course other diagrammatic contributions.  There are terms where we make other contractions such as shown in Fig.~\ref{figd4}.  These diagrams are, however, lower order in $N_a$, i.e., higher order in $N_a^{-1}$ (this is one reason why we introduced $a$ to help reduce the number of such diagrams; a more important reason is below).  Even if these diagrams were not lower order, these diagrams also have an uncontrolled sign.

\begin{figure}
\includegraphics[width=2.5in]{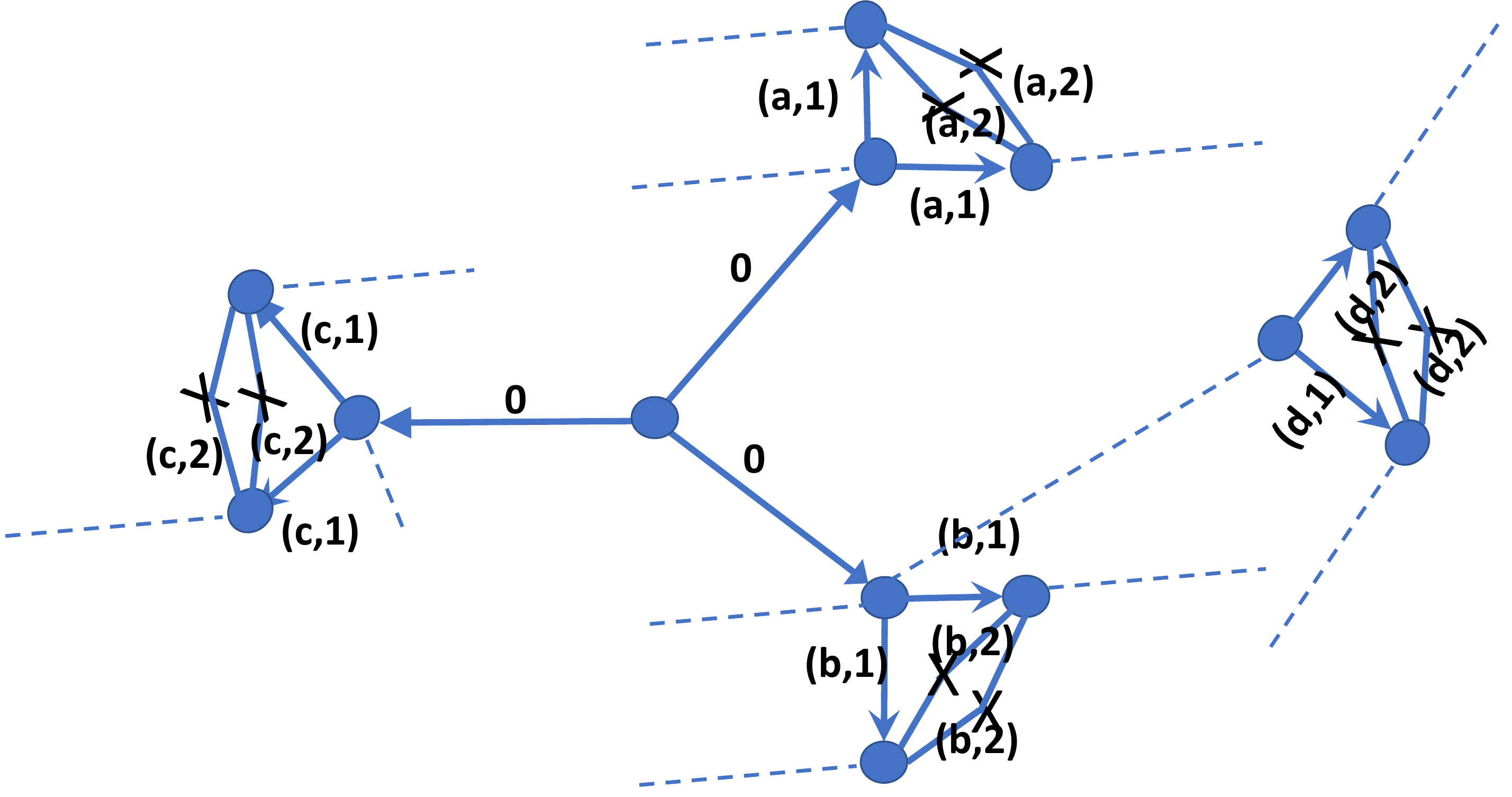}
\caption{Subleading diagrammatic contribution: higher order in $D^{-1}$.}
\label{figd4a}
\end{figure}

\begin{figure}[!htb]
\includegraphics[width=2.5in]{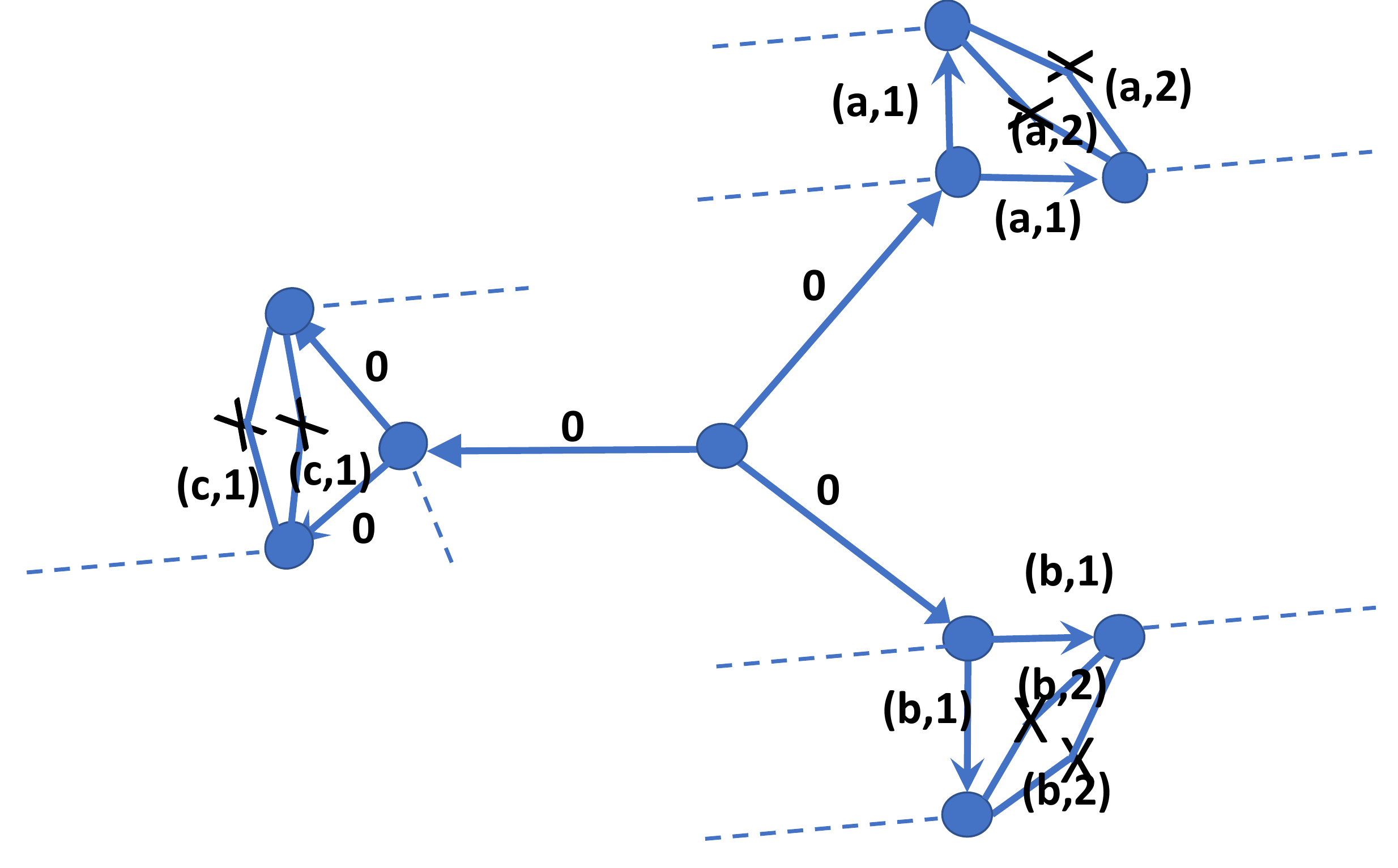}
\caption{Subleading diagrammatic contribution: same order in $N_a$ but lower order in $N_p$.}
\label{figd5}
\end{figure}

Another possibility is to take a diagram with fewer vertices but such that one still has a trivalent graph after removing dashed lines: this will be lower order in $D^{-1}$ for $J\Phi>>D^{-3/4}$.

Yet another possibility is to consider higher order expansion of the sine, rather than approximating $\sin(\phi J_{i,i+1} \ze_{i+1}^2) \approx \phi J_{i,i+1} \ze_{i+1}^2$.
That is, in a given step of the QAOA we may commute a term with $H$ multiple times, rather than just once.  
Diagrammatically, we will represent this by allowing dashed line leaving a vertices to connect to further trivalent vertices, representing the commutator of some $\ex$ with terms in $H$.  However, if we define a graph by removing external dashed lines (but retaining dashed lines connecting a pair of vertices), we find that now not every edge
comes with a factor of $\Theta(\sqrt{D})$ because not every edge of the graph has an associated external dashed line that we removed.
This reduces the order of the graph in $D$ for a given number of edges.  We expect that by adjusting $J$, we can make such diagrams negligible in $D^{-1}$ compared to those in Fig.~\ref{figd3}.  See for example the diagram of Fig.~\ref{figd4a}.  This has added $3$ additional vertices compared to that of Fig.~\ref{figd3}, but only added $4$ factors of $D^{1/2}$, rather than $3 \cdot (3/2)$ factors.  Hence, so long as $(J\Phi)^3 D^2 <<1$, this diagram is negligible.  So, in the regime $D^{-3/4} << J\Phi <<D^{-2/3}$, we can make this diagram negligible.  
 For $\Phi>>D^{-2/3}$, we can reduce $J$ to enter this regime, so long as $\Phi$ is not too large.
We expect that other such diagrams can be made negligible similarly. 

A final possibility (and the most important one to consider) is terms which involve $J_0$ beyond first order.  For example, see the diagram of Fig.~\ref{figd5}.
This, however, is lower order in $N_p$.  
Indeed, the only diagram that is leading order in $D^{-1}$, and is leading order in $N_a^{-1},N_p^{-1}$ for the given order in $D^{-1}$, is that in Fig.~\ref{figd3},
and since $J_0$ appears once in this diagram, the sign of this diagram cannot be controlled: it could lead to a positive or negative contribution to the expectation value of $H$.
To see that this is the only such diagram, note that for the given number of vertices, the edges labelled $l+1$ form a closed loop.  We must maximize the number of such closed loops.

\section{Discussion}
We have shown several specific results, such as that the QAOA is likely to be limited in its performance on bounded degree MAX-3-LIN-2, no matter how many steps one considers, and that local classical algorithms can outperform the QAOA in many cases.  More generally, however, the results here point to a pattern.  It seems like the QAOA is simply a way of writing a quantum circuit which does something similar to a very simple classical algorithm: initialize spins in some configuration and then update spins depending on those around it.
For a one-step version of the algorithm, the update is usually very simple: if the spins around it apply a strong enough force to the given spin, then it changes in a way to increase the objective function.  To the extent that the QAOA works, it seems that it works because the classical algorithm works; there does not seem to be any general explanation in the literature as to {\it why} such a quantum circuit should be better than such a classical algorithm.

While one argument given for the QAOA is that (roughly) ``with a sufficiently large number of steps, the algorithm converges to the adiabatic algorithm, so it must succeed in producing the optimal state in that limit, and so it will be useful for a small number of steps", this same argument can be applied to the local tensor algorithms here.  They can reproduce simulated annealing which also finds the exact ground state with a sufficiently slow anneal.  Alternatively, the generalization can also reproduce any diagrammatic sum, again allowing exact optimization in this limit.
Further, these kinds of arguments do not in themselves seem to be sufficient as they describe only some limiting behavior where a certain parameter becomes large.  As an example of a case where similar arguments are made, matrix product states are very useful for practical studies on one-dimensional quantum systems.  However, while it is true that for large enough bond dimension any state can be described as a matrix product state, the real reason that these states are interesting is that many practical important states can be represented to high accuracy as matrix product states with a low bond dimension (see \ocite{White1992,Vidal2003,Hastings2007,Pirvu2010} among many other works).

Thus, it is not apparent why one should believe that the QAOA will be better than local classical algorithms.  Numerical extrapolation of the performance of the QAOA from small sizes as can be currently simulated is difficult; and the QAOA has a number of adjustable parameters, especially as the number of steps is increased, and one should take care that the number of parameters not become large compared to the problem size.  However, on large systems, it seems like current classical algorithms are better than the QAOA, even including local tensor algorithms such as those studied here.
Thus, it seems that one should investigate further local classical algorithms such as those here which depend upon adjustable parameters; this is likely to be at least as promising as the QAOA.  

We have also discussed the possibility of more general local quantum/classical algorithms.
Note, for example, that we obtained an improved performance for $D=3,4$ by allowing the entries of the vector to be chosen from a continuous distribution rather than from $\{-1,+1\}$; that is, we chose the degrees of freedom associated with each variable from a larger set.  Similarly, one might consider a quantum algorithm using  local degrees of freedom, both classical and quantum (with the quantum degrees of freedom in a Hilbert space of much larger dimension than the number of choices of the local variable) which uses some arbitrary quantum channels to communicate between degrees of freedom on neighboring spins and to update the degrees of freedom in some general way (this would not be a QAOA but it would still be a local algorithm). 
 It has not been ruled out that such an algorithm would yield an improvement over a local classical algorithm, though it seems unlikely that it would yield much, if any, improvement for approximate combinatorial optimization problems.
At  the same time, one should expect that  existing classical algorithms which allow nonlocal communication and which allow depth to depend on problem size will generally outperform local algorithms, both quantum and classical.

Finally, it is interesting to consider the number of parameters that must be tuned in these algorithms.  The one-step QAOA involves two real parameters which must be optimized, though one of these parameters takes the same value in many applications so more fairly one may say that the one-step QAOA requires tuning a single real parameter.  At the same time, the threshold algorithm requires tuning only a single discrete parameter.  For the cases $D=6,11$ we tuned one discrete parameter and one real parameter, while for $D=3,4$ we tuned only a single real parameter.  We can also consider the amount of communication required.  The one-step QAOA requires communicating quantum information between different spins, while the threshold algorithm requires communicating only a single bit from one spin to its neighbors and the generalization in the case of $D=3,4$ requires communicating one real number.

{\it Acknowledgments---} I thank J. Haah, S. Jordan, and D. Wecker for useful discussions.

\section{Numerical Results for Triangle Free MAX-CUT}
\label{numres}

In this section we briefly list some of the numerical results for triangle-free MAX-CUT.  The performance numbers for the QAOA are obtained from the exact formula for the optimal choice of parameters\ocite{wang2018quantum,ryan2018quantum}.  See table \ref{tx}.

\begin{table}[!htb]
\begin{tabular}{l|l|l|l}
D & Threshold Algorithm & Threshold Value& Quantum\\
\hline
2 & 0.2500 & 2 & 0.2500 \\
3 & 0.1875 & 3 & 0.1925 \\
4 & 0.1406 & 3 & 0.1624 \\
5 & 0.1562 & 4 & 0.1431 \\
6 & 0.1221 & 5 & 0.1294 \\
7 & 0.1282 & 5 & 0.1190 \\
8 & 0.1166 & 6 & 0.1108 \\
9 & 0.1077 & 6 & 0.1040 \\
10 & 0.1077 & 7 & 0.0984 \\
11 & 0.0925 & 7 & 0.0936 \\
12 & 0.0987 & 8 & 0.0894 \\
13 & 0.0886 & 9 & 0.0858 \\
14 & 0.0905 & 9 & 0.0825 \\
15 & 0.0853 & 10 & 0.0796 \\
16 & 0.0833 & 10 & 0.0770 \\
17 & 0.0816 & 11 & 0.0747 \\
18 & 0.0771 & 11 & 0.0725 \\
19 & 0.0778 & 12 & 0.0705
\end{tabular}
\caption{Improvement over random assignment of algorithms for triangle-free MAX-CUT.  The degree $D$ is shown in the first column.  ``Threshold Algorithm" denotes the algorithm with a threshold, while ``Threshold Value" denotes the optimal value of the threshold.  ``Quantum" denotes the performance of the one-step QAOA.  The threshold algorithm is better for all $D$ in this range except for $D=3,4,6,11$, as well as for all $D$ up to $1000$ as shown numerically.}
\label{tx}
\end{table}
\bibliography{lim-ref}
\end{document}